# Confusion in Thermodynamics.

J. Dunning-Davies and D. Sands,
Physics Department,
Hull University,
Hull HU6 7RX,
England.

email: j.dunning-davies@hull.ac.uk
d.sands@hull.ac.uk

**Abstract.**

For a long time now, confusion has existed in the minds of many over the meaning of various concepts in thermodynamics. Recently, this point has been brought to people's attention by two articles appearing on the well-known archive (arxiv) web site. The content of these two pieces serves to illustrate many of the problems and has occasioned the construction of this answer to at least some of them. The position of the axiom proposed by Carathéodory is central in this matter and here its position is clarified and secured within the framework of thermodynamics. In particular, its relation to the First Law is examined and justified.

**Introduction.**

There is little doubt that, although based on phenomena and, possibly more importantly, experiences with which virtually everyone is familiar, confusion does arise in the minds of many when it comes to understanding thermodynamics. However, as was stated on the dust cover of Landsberg's first book on the subject [1],

*'Thermodynamics is among the most abstract branches of physics'.*

Considering the beginnings of the subject were so firmly rooted in so seemingly practical a subject as the theory of heat engines, this appears an almost astonishing statement but a little delving into the theory shows it to be a fairly accurate assessment of the subject and probably indicates one of the reasons for the confusion arising in so many minds. Added to this thought, it is often forgotten that, when considering theories of physics in general, the discussion concerns theories which purport to describe perceived physical happenings. As such, it should be remembered always that it is the *physics* which is all-important; it is the physical observations which must always provide the impetus in any subsequent investigations. In all this, the mathematics is merely a tool; a very important tool, but still a tool. Therefore, any deductions made must be checked against observed physical fact. It is the *physics* which must dominate! On the other hand, it often proves difficult to do this as the mathematical theory often appears abstruse to the uninitiated and frequently serves to add to the mysticism some might wish to attach to their chosen field. There is little doubt that this is one factor leading to confusion in the minds of many in various areas of science, of which thermodynamics is but one.

**The Carathéodory and Kelvin Forms of the Second Law.**

The argument employed in a recent posting [2] on the well-known Cornell University administered archive site by Radhakrishnamurty is a perfect example of where this confusion has helped towards a chaotic misunderstanding of some of the basics of thermodynamics. The claims concerned the validity of the proofs of the equivalence of the Carathéodory and Kelvin forms of the Second Law of thermodynamics. He based his claim on the version of the Second Law which appears in Fermi's book on thermodynamics [3]. As printed, this states that

*"A transformation whose only final result is to transform into work heat extracted from a source which is at the same temperature throughout is impossible."*

As with all forms of the Second Law, whether old or more modern, it is important to read *all* the words and realise that they *all* form crucial parts of the statement. In the version quoted here, one highly important part is '*whose only final result*'. This means quite simply that, at the end of the procedure, the only end result is the transformation of an amount of heat into work. This means that, at the end of the procedure, the entire system must be back where it started or, in other words, the whole process must have been cyclic. Nowadays, statements of the law usually refer specifically to everything occurring in a cycle. However, in Kelvin's day, as is seen quite clearly from the writings of Tait [4], Kelvin was constantly discussing cycles and possibly saw little need to include the word explicitly in his statement of the Second Law. The fact that cycles were always under consideration follows from the fact that thermodynamics arose out of the study of heat engines which were so important for raising and lowering cages in mines and for pumping water out of many of those mines. Such engines work in cycles.

What Landsberg showed in his article [5] was that, if Carathéodory's accessibility criterion was untrue, then an amount of heat could be converted completely into work in a cycle, contrary to the Kelvin form of the Second Law. In Landsberg's cycle, although not stated explicitly since such a statement was not necessary, heat is given to the system from a heat reservoir and, therefore, that heat is transferred from a source at constant temperature. The cycle is completed by returning to the starting point by using an adiabatic process as allowed if Carathéodory's axiom is not valid since, if not valid, the said axiom would imply that, in the neighbourhood of a state, *all* other states would be adiabatically accessible from it. Hence, the proof that Kelvin's form of the Second Law implies the validity of Carathéodory's axiom is valid.

Of course, some may point out that one of the processes here is isothermal and so occurs at a constant temperature, whereas the other is adiabatic and so is accompanied by a temperature change. Hence, the above mentioned cycle is impossible and so the proof invalid. This is simply not true. The argument advanced by Landsberg simply makes no assumptions about isothermal and adiabatic processes except that, in the first, heat is given to the system at a fixed temperature while, in the second, work is done but no heat is transferred. The Second Law in its more traditional form due to Kelvin then rules out the possibility of such a cycle and the validity of the Carathéodory axiom is verified. There is no contradiction here.

A similar argument is used to see in what sense the reverse is true [6] and it is found that Carathéodory's axiom is more general than the usual form of Kelvin's statement of the Second Law but equivalent to the modified form applicable when negative temperatures are considered.

Hence, the claims of Radhakrishnamurty are unfounded and, if you consider the form of Kelvin's statement usually in use today [7] and which was the one used by both Landsberg and Dunning-Davies in their discussions:

"It is impossible to transform an amount of heat completely into
work in a cyclic process in the absence of other effects",

you can see immediately that there are no real differences between it and the form Fermi attributes to Kelvin; the main difference is that this later version specifically mentions cycles, while the concept is implied in the quoted Fermi version. This, of course, is yet another point that Radhakrishnamurty fails to appreciate; he constantly criticises Landsberg for using an 'incorrect' form of Kelvin's statement of the Second Law. This situation illustrates, once again, how confusion and lack of understanding can creep into people's knowledge of thermodynamics. However, in this particular instance, it is due to a lack of care in reading the separate statements of – in this case – the Second Law. It is always vitally important to read and digest each and every word of such statements. As pointed out earlier, in the form attributed by Fermi to Kelvin, the complete meaning of the words '*whose only final result*' must be absorbed and followed. The more modern wording, given above and used by Landsberg, merely singles out for special mention the fact that everything must occur in a cycle so that, after everything, the entire system ends up in exactly the same position from which it started. Not realising that both forms are actually identical indicates a lack of basic understanding and an inability to read and absorb information in totality.

**Carathéodory and the First Law.**

Admittedly, as acknowledged earlier, thermodynamics is a subject which causes problems for people at all levels of academia. This problem has begun to be addressed [8] in a variety of ways but, from the outset, it must be acknowledged that one way in which problems are created is through unclear and muddled thinking and writing. This is a problem which has plagued thermodynamics almost from the very birth of the subject. The writings of Clausius are a typical case in point and it might be argued that the foundations of much of the confusion existing today rest in his early articles, although here attention is focused on problems associated with the more modern approach associated with the name of Carathéodory. The note to which reference is being made here [9] is an excellent example of the problem.

Firstly, while Carathéodory was invited to make a contribution to the subject by Max Born and while Born himself made contributions, notably via his book *Natural Philosophy of Cause and Chance* (Dover, New York; 1964), the new more abstract approach was clarified and modified by several people, mainly P.T. Landsberg in England, L. A. Turner, F. W. Sears and M. W. Zemansky in the U.S.A., and H. A. Buchdahl in Australia. In fact, the end result of their work has seen the virtual abandonment of Carathéodory's axiom from use but the retention of methods originally associated with it, namely the technique for deriving the existence of both an absolute temperature and entropy from the Second Law, as well as the equation representing the Second Law:

$$d'Q = TdS \qquad (1)$$

where the ´ indicates an inexact differential. Hence, the first problem with the cited article is that it doesn't even represent the historical facts too accurately.

However, in this second article by Radhakrishnamurty [9], it is noted that, if a system undergoes an adiabatic change between two equilibrium states $A$ and $B$, the First Law reduces to

$$dU = d'W \qquad (2)$$

That is, during the change, work done produces a change in internal energy.

The author then claims that 'state $B$ is not reachable by an adiabatic process from state $A$ only if $dU \neq d´W$.' He goes on to claim that 'in other words state $B$ is not reachable from state $A$ by an adiabatic process only if the First Law is violated.'

It is the second sentence here which is totally misleading. However, not only is the conclusion misleading, but one must also question the methodology involved. Adiabatic inaccessibility means that there is no work process that leads to the difference in energy between two states, and one way to demonstrate the validity or otherwise of this assertion is to show by logical argument that the assumption of such a process leads to a contradiction. For example, suppose that Carathéodory is invalid and that all states in the neighbourhood of every other state are accessible by an adiabatic process. Thus, state $A$ is accessible from $B$ and vice versa. This would mean that all adiabatic processes are reversible and that in consequence there is no such thing as an irreversible adiabatic process and would render much of the development of thermodynamics invalid. If we consider by way of example the free expansion from, say $A$ to $B$, we know that state $A$ cannot be attained from $B$ by an adiabatic process as the volume must be decreased and work done on the gas. It is necessary, therefore to extract heat from the system in order to return to state $A$. Clausius' theorem that, around a closed, irreversible cycle the integral of $dQ/T < 0$, leads to the same result. Clausius' theorem, therefore, implies Carathéodory but, more importantly, we have shown that the initial assumption leads to a contradiction, namely that there are such things as irreversible adiabatic processes, the free expansion being but one example.

Radhakrishnamurty seems to believe he has shown, by a similar process, that the assumption that Carathéodory is valid leads to a violation of the First Law. On the contrary, he has assumed an adiabatic process between two states and then simply considered the case when

$$dU \neq d'W \qquad (3)$$

This is nothing more than a contrary assumption to the original, namely that the process linking the two states is not adiabatic. By the First Law, which can be written generally as

$$d'Q = dU - d'W, \qquad (4)$$

it is seen that, if a non-adiabatic process is being considered, the term $d´Q$ makes a contribution or, in other words, the difference between the terms $dU$ and $d´W$ is made up by a quantity of heat $d´Q$. Hence, the conclusion supposedly deduced from the first quoted sentence from quoted article [9] simply does not follow and Carathéodory's axiom is perfectly compatible with the First Law of Thermodynamics which is, in fact, what Radhakrishnamurty has actually demonstrated.